\documentclass[12pt]{article}
\usepackage{helvet}
\usepackage{color}
\usepackage{fancyhdr}
\usepackage{float}
\usepackage{vmargin}
\usepackage{amsmath,amsfonts,amssymb}
\usepackage{mathtools}
\usepackage{graphicx}
\usepackage{lastpage}
\usepackage{here}
\usepackage{tabularx}
\usepackage{tabulary}
\usepackage{booktabs}
\usepackage{hyperref}
\usepackage{algorithm}
\usepackage{caption}
\usepackage{subcaption}
\usepackage[noend]{algpseudocode}
\usepackage[table]{xcolor}
\usepackage[comma,numbers]{natbib}
\usepackage{graphicx}
\usepackage[utf8]{inputenc}

\newcommand{\beginsupplement}{%
        \setcounter{section}{1}
        \setcounter{subsection}{0}
        \renewcommand{\thesubsection}{S.\arabic{subsection}}%
        \setcounter{page}{1}
        \setcounter{table}{0}
        \renewcommand{\thetable}{S\arabic{table}}%
        \setcounter{figure}{0}
        \renewcommand{\thefigure}{S\arabic{figure}}%
        \setcounter{algorithm}{0}
        \renewcommand{\thealgorithm}{S\arabic{algorithm}}%
     }

\begin{document}
 
 \begin{Large}
\noindent \textbf{INtERAcT: Interaction Network Inference from Vector Representations of Words}
\end{Large}\\
\begin{footnotesize}
\noindent Matteo Manica\textsuperscript{1,2,*}, Roland Mathis\textsuperscript{1,*}, María Rodríguez Martínez\textsuperscript{1, $\dagger$}\\
\noindent \textit{ \{tte,lth,mrm\}@zurich.ibm.com}\\
\textsuperscript{1} IBM Research Zürich\\
\textsuperscript{2}	ETH - Zürich\\ 
\textsuperscript{*} Shared first authorship \\
\textsuperscript{ $\dagger$} Corresponding author
\end{footnotesize}

\section*{Abstract}

\subsection*{Motivation}
In recent years, the number of biomedical publications has steadfastly grown, resulting in a rich source of untapped new knowledge. Most biomedical facts are however not readily available, but buried in the form of unstructured text, and hence their exploitation requires the time-consuming manual curation of published articles. Here we present INtERAcT, a novel approach to extract protein-protein interactions from a corpus of biomedical articles related to a broad range of  scientific domains in a completely unsupervised way. INtERAcT exploits vector representation of words, computed on a corpus of domain specific knowledge, and implements a new metric that estimates an interaction score between two molecules in the space where the corresponding words are embedded. 

\subsection*{Results}

We demonstrate the power of INtERAcT by reconstructing the molecular pathways associated to 10 different cancer types using a corpus of disease-specific articles for each cancer type. We evaluate INtERAcT using STRING database as a benchmark, and show that our metric outperforms currently adopted approaches for similarity computation at the task of identifying known molecular interactions in all studied cancer types. Furthermore, our approach does not require text annotation, manual curation or the definition of semantic rules based on expert knowledge, and hence it can be easily and efficiently applied to different scientific domains. Our findings suggest that INtERAcT may increase our capability to summarize the understanding of a specific disease using the published literature in an automated and completely unsupervised fashion. 

\subsection*{Contact}

mrm@zurich.ibm.com or tte@zurich.ibm.com.

\vspace{1\baselineskip}

\noindent \textbf{Keywords:} molecular networks, natural language processing, text mining, deep learning, word embeddings, cancer.

\section{Introduction}

As the number of scientific publications continues to grow exponentially, search engines such as PubMed\footnote{\url{https://www.ncbi.nlm.nih.gov/pubmed/}} provide an unprecedented amount of information in the form of unstructured written language. With the accelerating growth of available knowledge -- particularly in the biomedical literature --  and the breakdown of disciplinary boundaries, it becomes unfeasible to manually track all new relevant discoveries, even on specialized topics. As an example, recent advances in high throughput experimental technologies have yielded extensive new knowledge about molecular interactions in the cell; however most of this knowledge is still buried in the form of unstructured textual information only available as written articles. 

As of October 2017, PubMed  comprises more than 27.8 million references\footnote{The current size of the database can be obtained by typing "1800:2100[dp]" into the search bar.} consisting of biomedical literature from MEDLINE, life science journals, and online books. Most references  include links to full--text content from PubMed Central\textsuperscript{\textregistered} (PMC\footnote{\url{https://www.ncbi.nlm.nih.gov/pmc/}}) -- a free full--text archive of biomedical and life sciences journal literature -- or publisher web sites. Currently 14.2 million PubMed articles have links to full--text articles, 4.2 million of which are freely available.
The numbers remain high even when focusing on specific fields such as prostate--cancer. For instance, a simple query\footnote{\url{https://www.ncbi.nlm.nih.gov/pmc/?term="prostate+cancer"}} for prostate--cancer related papers 
on PMC returns 143321 publications\footnote{Number obtained as of 12 October 2017}.
While a fraction of the information currently available in biomedical publications can be extracted from public databases, the rate at which new research articles are published greatly exceeds the rate at which this information can be currently processed, resulting in an ever wider gap between available knowledge and easily accessible information, e.g. information stored in a database. 
Clearly the development of novel methodologies  that can automatically analyze textual sources, extract facts and knowledge,  and produce summarized  representations that capture the most relevant information are needed more than ever.

We present here a novel approach to automatically extract knowledge from biomedical publications. Specifically, we  focus on the problem of identifying and extracting Protein–Protein  Interactions  (PPIs)  from a disease--specific text corpus and building an interaction network. While our approach is generic and can be applied to any knowledge domain, we demonstrate its strength using the biomedical literature related to prostate--cancer (PC), a complex disease with multi--factorial etiology.
PC is the second most common cancer type and the fourth leading cause of cancer death in men worldwide~\cite{ferlay_cancer_2015}. Despite the large number of newly diagnosed cases, the majority of cases in older men are clinically insignificant, meaning that  the life expectancy of the patient is shorter than the time required by the disease to manifest any symptoms~\cite{Zlotta17072013}. However a small fraction of new cases are aggressive cancers that require intervention. The current prognostic factors are not sufficient to precisely stratify these two groups~\cite{Cooperberg16062009}, and thus PC is prone to overdiagnosis and treatment associated with debilitating side effects\cite{Chou2011}. 

While various approaches to automatically extract PPIs  information  from  unstructured text  are already available, many of these  methods  require feature engineering and expert-domain knowledge for good performance, hence preventing full automation. Commonly proposed methodologies  exploit machine learning approaches~\cite{tikk_comprehensive_2010,tjioe_discovering_2010}, data mining tools~\cite{mandloi_palm-ist:_2015}, co-occurrences~\cite{barbosa-silva_pescador_2011,fleuren_identification_2013,raja_ppinterfindermining_2013,usie_biblio-metres_2014}, or rules--based text mining~\cite{torii_rlims-p_2015}.

Recently, word embedding techniques based on deep learning have been proposed as a more advanced approach to process textual information in an unsupervised fashion. 
Word embedding is a term used to identify a set of methods for language modelling and feature learning,  where  words  in  a  vocabulary are  mapped  into  vectors  in  a  continuous, high  dimensional space, typically of several hundred dimensions~\cite{collobert_unified_2008}. In this representation, words that share a similar context in the corpus are located in close proximity in the word embedding vector space.
Besides representing  words'  distributional  characteristics, word--vectors can capture the semantic  and  sequential  information  of  a  word  in a text, providing a richer vector representation than frequency--based approaches. 
Word--vector representations have gained broad  recognition  thanks  to  the  recent  work  of  Mikolov  et  al.~\cite{mikolov_efficient_2013,mikolov_distributed_2013},  who demonstrated that word embeddings can facilitate very  efficient  estimations  of  continuous--space  word representations from huge datasets ($\sim$ 1.6 billion words).

Since this seminal work, word embeddings based on neural networks have been used to address different tasks of natural language processing. For instance, word embeddings were used in~\cite{nie_embedding_2015} for the task of event trigger identification, i.e. to automatically detect words or phrases that typically signify the occurrence of an event. In~\cite{zhou_event_2014}, a combination of features extracted from a word embedding plus syntactic and semantic context features was used to train a support vector machine classifier for the task of identifying event triggers. Such approaches have been shown to be efficient in identifying the semantic and syntactic information of a word and incorporate it into a predictive model. 
Word embeddings have also been used as token features -- semantic units of words and characters extracted from a corpus for further processing -- to extract complete events represented by  their trigger words and  associated arguments ~\cite{li_using_2015};
to build knowledge regularized word representation models that  incorporate  prior knowledge into distributed word representations for semantic relatedness ranking tasks~\cite{wang_incorporating_2015}; 
and to simultaneously analyze the semantic  and contextual relationship between words~\cite{jiang_unsupervised_2016}.
Finally, alternative deep learning approaches based on autoencoders and a deep multilayer neural network have been used to extract PPIs, where the features were extracted by a Named  Entity Recognition module coupled to a parser and principal component analysis~\cite{zhao_protein-protein_2016}.

While these approaches have shown the versatility of word embeddings to support text analysis through current natural language processing  tools, approaches that can automatically extract molecular interactions from unstructured text in a completely unsupervised manner are still missing. To bridge this gap we present our methodology hereby referred as INtERAcT (Interaction Network infErence from vectoR representATions of words).
Our approach can be summarized as follows. We first create a word embedding from a corpus of freely available publications related to prostate--cancer. 
Next, we cluster the learned word--vectors in the embedded word--space and find groups of words that convey a close semantic and contextual similarity.
Then we  focus on proteins and predict PPIs using a novel similarity measure based on the Jensen--Shannon divergence.  To demonstrate the generalization potential of our approach to other domains of knowledge, we repeat the exercise and apply INtERAcT to a corpus of publications related to 10 different cancer types, and validate our results using STRING\footnote{\url{https://string-db.org/}} database~\cite{szklarczyk_string_2015}  as a benchmark.

\section{Results}

\subsection{Applying INtERAcT on prostate--cancer publications}
\label{subsection:prostate}

{ \bf Building a word embedding specific to prostate--cancer:} In the following section we describe the application of INtERAcT to the problem of reconstructing a prostate--cancer pathway. Text pre-processing and building of the word embedding follows the methodology described in Section~\ref{sec:methods}. 
Briefly, a text corpus is assembled by downloading the XML version of $\sim$ 140000  PubMed Central publications matching the query {\it "prostate cancer"}. Only abstracts are processed, as they provide a  concise and clean summary of the  article's main findings. Rare words and bi--grams occurring less than 50 times in the corpus are removed.  The remaining sentences are tokenized -- segmented into linguistic units -- and used to build a word embedding.
After processing (see section \ref{sec:processing}), our dictionary is composed of $\sim$ 21000 single words and common bi--gram, e.g. prostate\_cancer, cell\_proliferation and gene\_expression. Using this dictionary, we build a word embedding using a vector representation of 500 dimensions and a context window of size 8 words (4 words to the right and the left of each target word). See 
Section~\ref{sec:methods} and Fig.~\ref{fig:skip_gram} for details.

\begin{figure}[!ht]
    \centering
    \includegraphics[width=0.9\columnwidth,
    clip]{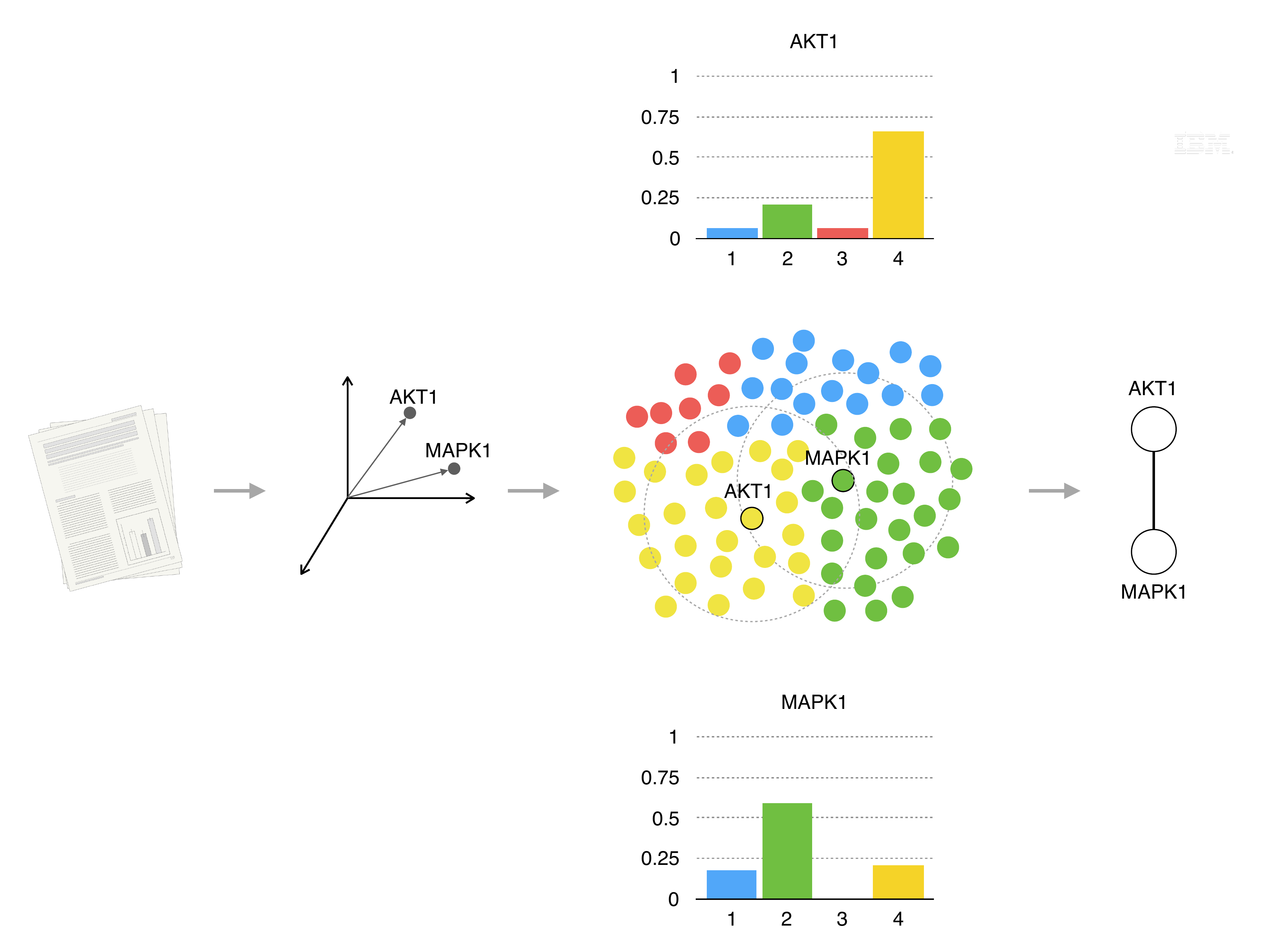}\\
    \caption{{\bf Schematic representation of INtERAcT.} The parsed text is used as input in a word embedding algorithm. The word vectors are clustered into groups of similar semantic meaning and the distributions of each word’s neighbors across clusters are used to compute and predict interactions between molecular entities.} 
    \label{fig:interact_graphical_abstract}
\end{figure}

\noindent {\bf Applying INtERAcT:} The results of the embedding are  clustered into  groups conveying similar semantic meaning using K--means with 500 clusters. We next identify the k-nearest neighbors of each protein as described in section~\ref{sec:methods}. The neighborhood size is set to  $k=$2000 in order to keep a balanced trade-off between number of neighbors and number of clusters. 
We use the cluster assignment distribution of selected words to calculate the pairwise similarity scores based on the Jensen–Shannon divergence (JSD) as shown in \ref{eq:similarity_score}. This last step is done on a subset of words -- in this example, a list of molecular entities defined using UniProt. We interpret this JSD-based distance metric as  the likelihood of a PPI. See Section~\ref{sec:methods} and Fig.~\ref{fig:interact_graphical_abstract} for details.

To benchmark the inferred network we focus on the list of molecular entities reported in the prostate--cancer pathway as defined by the Kyoto Encyclopedia of Genes and Genomes\footnote{\url{http://www.genome.jp/dbget-bin/www_bget?pathway+map05215}} (KEGG) and apply INtERAcT to the task of reconstructing the connectivity between   these entities.
Out of the 87 molecular entities that constitute the KEGG pathway, 67 are found in the embedding, and thus can be used as a validation set.

\begin{figure}[!ht]
    \centering
    \begin{subfigure}[b]{0.45\textwidth}
            \centering
              \includegraphics[width=\textwidth]{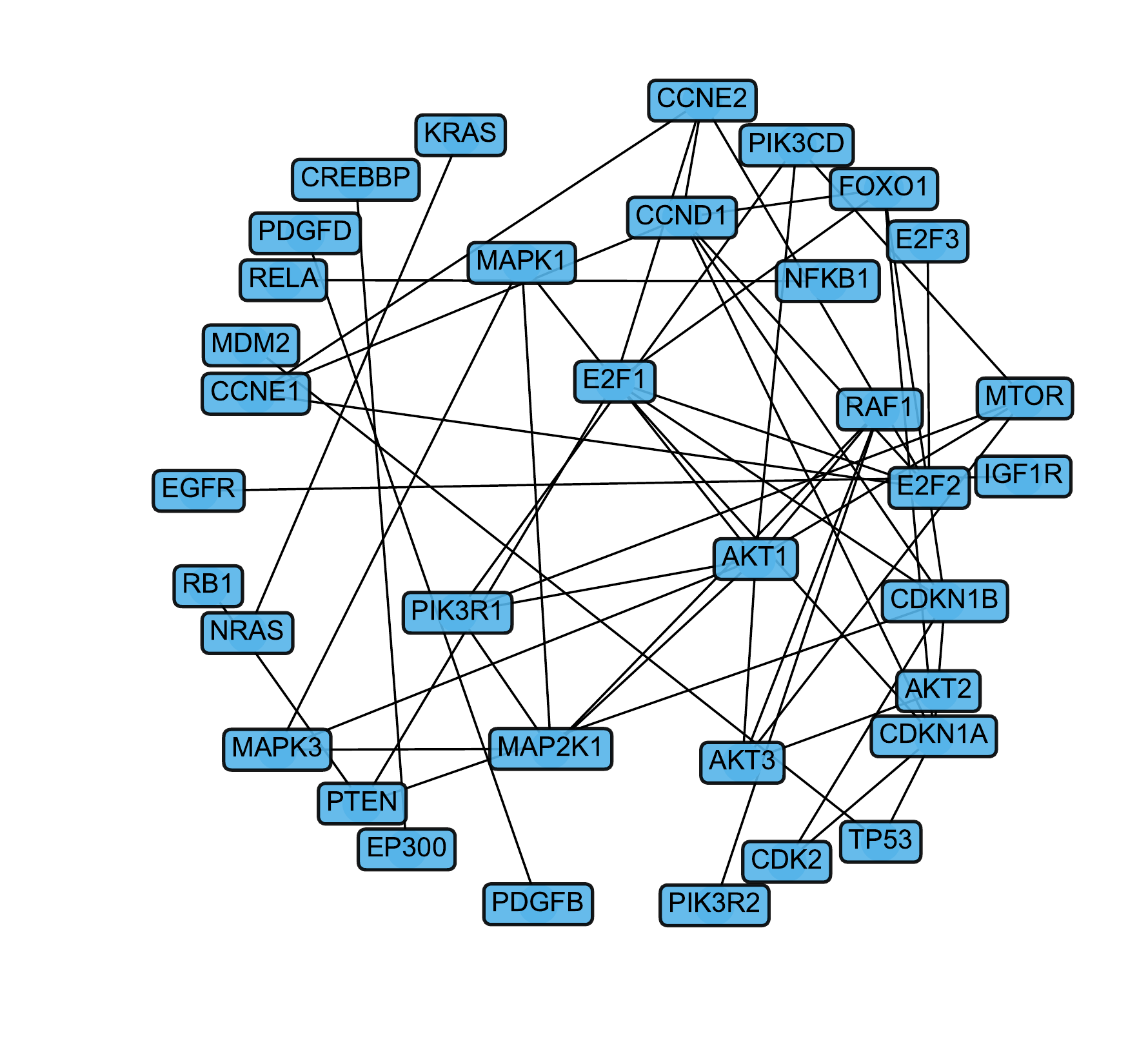}
            \caption{\label{fig:prostate_cancer_network_interact}}
    \end{subfigure}
    \begin{subfigure}[b]{0.45\textwidth}
            \centering
            \includegraphics[width=\textwidth]{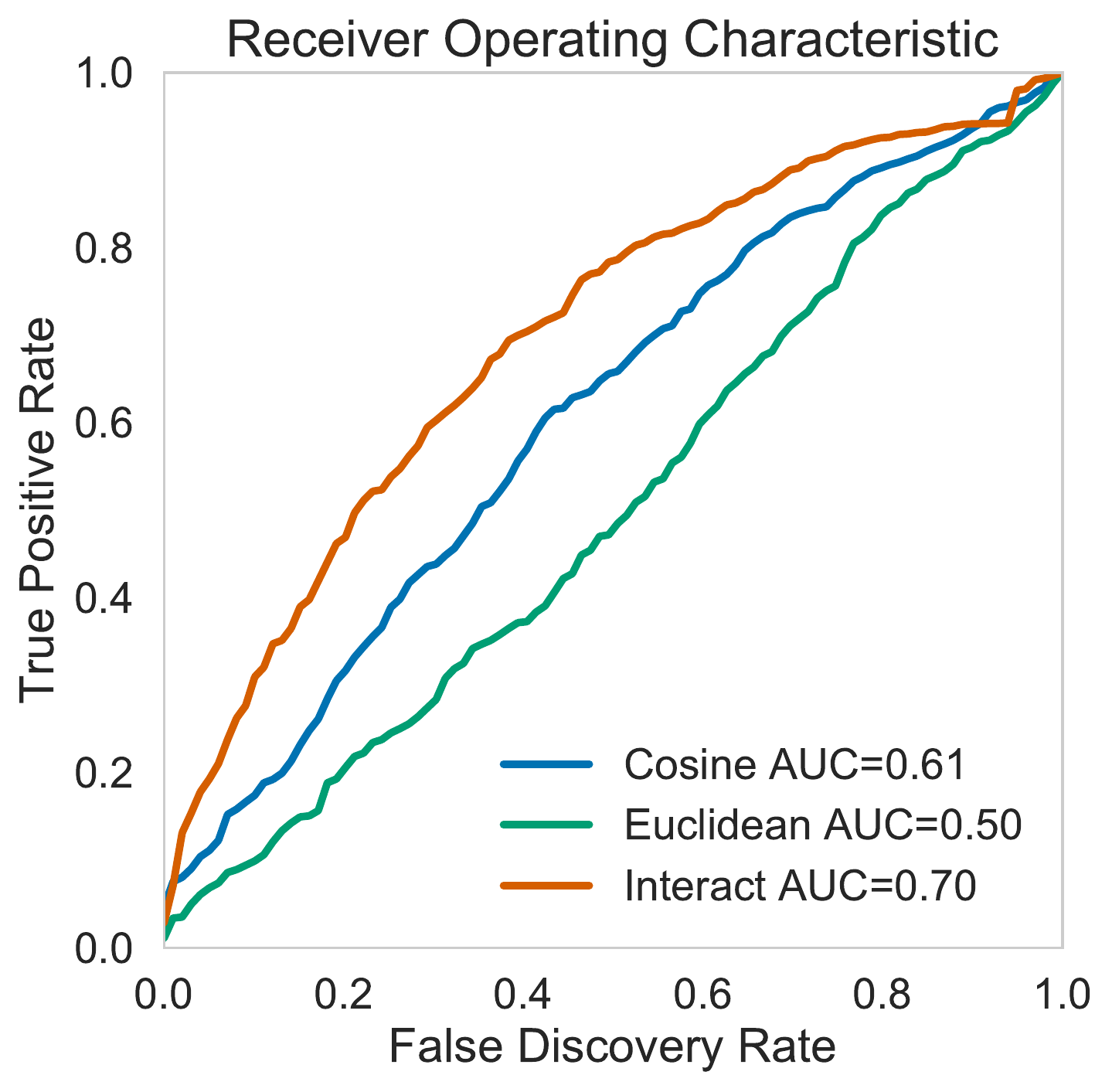}
            \caption{\label{fig:prostate_roc_plots}}
    \end{subfigure}
    \caption{\label{fig:prostate_results}
    {\bf (a) Top 50 prostate--cancer protein--protein interactions inferred by INtERAcT}.  The prostate--cancer gene set was defined according to the Kyoto Encyclopedia of Genes and Genomes (KEGG) prostate--cancer pathway, and includes molecular entities known to be important in prostate--cancer onset and development. The interactions and associated scores were computed using a word embedding trained on $\sim$140000 prostate--cancer abstracts freely available on  PubMed Central and INtERAcT, our proposed methodology to extract functional interactions from a word embedding. 
    {\bf (b)  Performance of INtERAcT on a prostate--cancer gene validation set compared to other distance measures using STRING as a ground truth.} We used a ROC (Receiver Operating Characteristic) curve to quantify the accuracy of the inferred interactions in a set of prostate--cancer-related genes. INtERAcT (orange curve) significantly outperforms alternative, commonly used metrics on a word embedding such as a cosine distance--based similarity (blue curve) and a similarity score based on the Euclidean distance (green curve). 
    } 
\end{figure}

We interrogate INtERAcT and query the interactions between the 67 proteins of our validation set. Fig.~\ref{fig:prostate_cancer_network_interact} graphically shows the top--50 inferred interactions in our prostate--cancer gene validation set. The full set of interactions with similarity scores can also be found as a table in the Supplementary Material~\ref{sup:prostate_scores}.
Please, notice that while KEGG provides a well-established reference for function-specific pathways, KEGG merges gene family members in a single node-entity (e.g. AKT1, AKT2 and AKT3 become AKT), and hence a direct comparison between KEGG prostate--cancer pathway and  INtERAcT inferred results is not possible.

\noindent { \bf Comparing INtERAcT to STRING:}
In order to assess the quality of our predictions, we  use  STRING\footnote{\url{https://string-db.org/}} database~\cite{szklarczyk_string_2015} as a benchmark. STRING is a comprehensive protein interaction database  currently including experimental data from DIP\footnote{\url{http://dip.doe-mbi.ucla.edu/dip/}}\cite{salwinski2004database}, BioGRID\footnote{\url{https://thebiogrid.org/}}\cite{chatr2017biogrid},  IntAct\footnote{\url{http://www.ebi.ac.uk/intact/}}\cite{intact_2004}, and  MINT\footnote{\url{http://mint.bio.uniroma2.it/}}\cite{licata2011mint}, and curated data from
BioCyc\footnote{\url{https://biocyc.org/}}\cite{caspi2007metacyc}, GO \footnote{\url{http://www.geneontology.org/}}\cite{ashburner2000gene}, KEGG\footnote{\url{http://www.kegg.jp/}}\cite{kegg_2000}\cite{kegg_2016}, and Reactome\footnote{\url{http://www.reactome.org/}}\cite{reactome_2014}. STRING provides a confidence score that integrates information about genomic proximity, gene fusion events, phylogenetic co--occurrences, homology, co--expression, experimental evidence of interaction, simultaneous annotation in databases and automatic text--mining \cite{franceschini_string_2013}.
Importantly for the sake of comparing STRING and INtERAcT results,  STRING text--mining is done by using a combination of co-occurrences and natural language processing based on a rule-based system~\cite{saric}.

To quantitatively evaluate  the goodness of INtERAcT predictions, we employ the Area Under a receiver operating characteristic Curve (AUC metric~\cite{florkowski_sensitivity_2008}) using STRING interactions as a ground truth, and compare our JSD-based score (Eq.~\ref{eq:similarity_score}) with other similarity scores commonly used in the literature, namely scores based on cosine and Euclidean distance. 
Figure \ref{fig:prostate_roc_plots} reports a summary of our findings. The Receiver Operating Characteristic (ROC) curve for the INtERAcT score (orange curve), a cosine--based distance score (blue line) and an Euclidean--based distance are comparatively shown. INtERAcT achieves a 0.70 AUC, significantly better than the cosine distance, which achieves a 0.61 AUC. The Euclidean based distance measure performs practically equivalent to a random predictor with an AUC value of 0.50. This poor performance is expected as the Euclidean distance, and more generically, $L_k$ norms, tend to map pairs of points to uniform distances in high dimensional spaces~\cite{aggarwal_surprising_2001}.
The curves' trends  reinforce the intuition that a neighborhood--aware metric is   better able to capture functional associations from unstructured text than methods that limit their analysis to the positions of word--vectors in the embedding. 
 
As an additional measure of agreement, we also compute the rank correlation between INtERAcT and STRING scores. To compute the correlation values, all predicted interactions by INtERAcT and STRING were used without applying any confidence cut-off. The INtERAcT and STRING scores (as downloaded on 19/10/17) used to compute the correlations are provided as additional supplementary tables. 
The resulting correlation value is positive and very significant ($\rho=0.31$, $p=1.6e^{-42}$), and is higher compared to the correlation obtained using cosine and Euclidean distance--based metrics ($\rho=0.29$, $p=4.1e^{-39}$ and $\rho=0.19$, $p=1.25e^{-16}$ respectively). 
INtERAcT outperforms again the cosine and Euclidian distance--based metrics. We note that while the correlation values obtained for INtERAcT and the cosine-based  scores seem to be relatively close, their difference turns out to be highly significant with a p-value of $p = 1.99 e^{-08}$ when the  number of interaction scores used to compute the correlations is taken into account (number of interactions = 132357). The significance of the difference of two correlation values can be computed using the Fisher z-transformation ~\cite{Fisher1915}, which  transforms the Spearman correlation values into  normally distributed variables whose difference can be evaluted used a standard t-test.

\subsection{Applying INtERAcT on other cancer pathways}
\label{subsection:pan_cancer}

\begin{figure}[ht]
    \centering
    \includegraphics[width=0.5\columnwidth,
    clip]{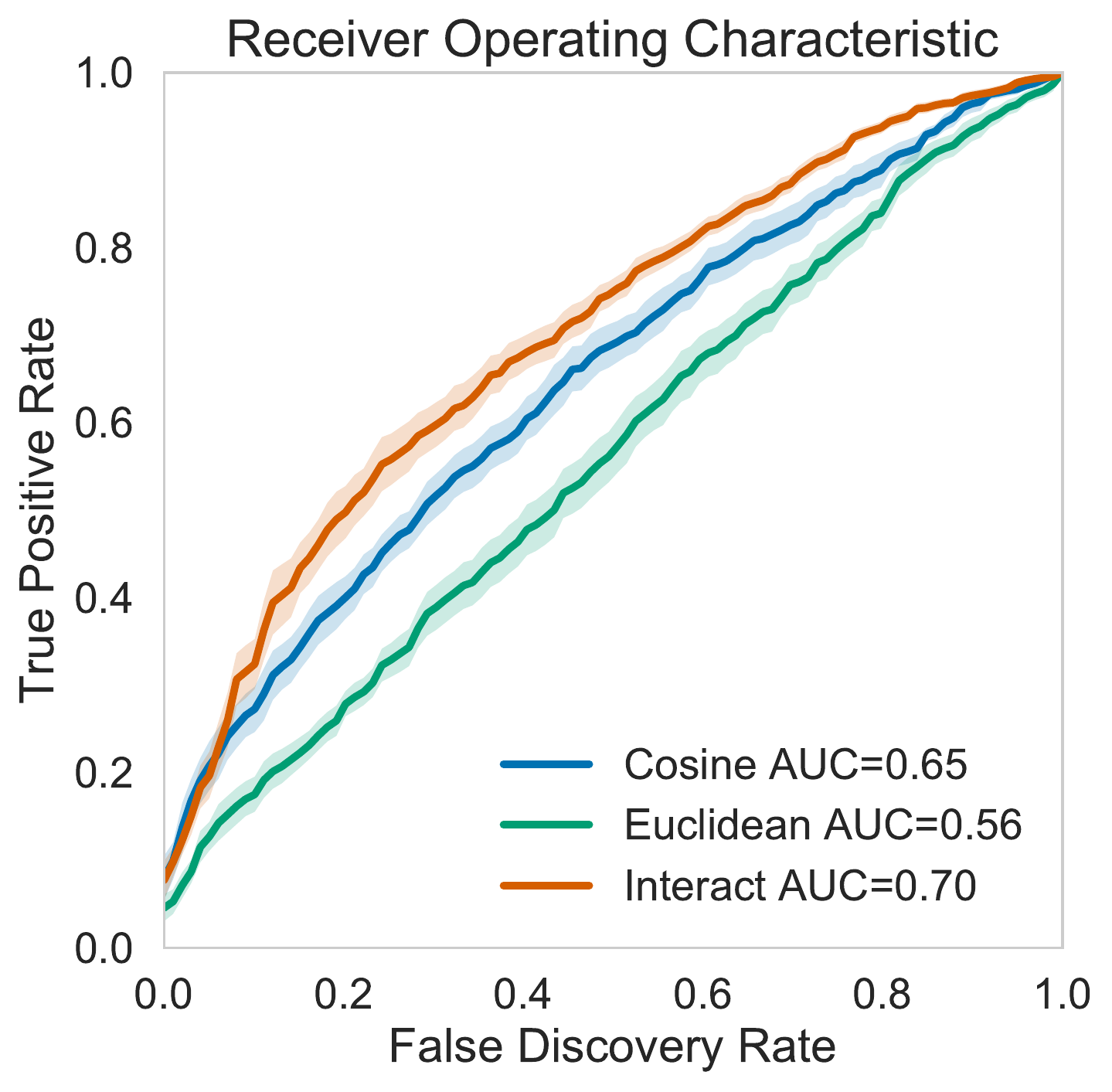}\\
    \caption{\label{fig:roc_plots}
    {\bf INtERAcT performance compared to other distance measures using STRING as a ground truth.} We use ROC (Receiver Operating Characteristic) curves  to quantify the quality and performance of  inferred interactions. The curves here reported refer to the inference performed on the KEGG cancer pathways considered in the analysis. Using naive approaches such as a similarity based on the Euclidean distance (green curve) between word--vectors led to poor results. Other methods such as cosine--based similarity (blue curve) showed an improvement. INtERAcT (orange curve) achieved the best performance predicting interactions reported in STRING. 
    The confidence intervals (CIs) at level 68\% are reported (one standard deviation from the mean).
    To generate the empirical distribution we used sampling with replacement at different false positive rates of the true positive rates given by the different pathways. The confidence intervals reported are at level 68\% (one standard deviation from the mean)
    } 
\end{figure}

We next focus on investigating the generalization of INtERAcT to other knowledge domains. For this task,  we extend our analysis to nine additional cancer types: acute myeloid leukemia, bladder cancer,
chronic myeloid leukemia, colorectal cancer, glioma, small cell lung cancer, non--small cell lung cancer, pancreatic cancer and renal cell carcinoma. The gene sets for each cancer type are taken from their respective cancer--specific pathway as annotated in KEGG. These cancer types are selected according to two criteria: first, there is a cancer-specific KEGG pathway to define a gene set, and second,  we could retrieve at least  10000 cancer-specific publications in PubMed Central. The second criterion is needed in order to obtain a corpus size that guarantees a good reconstruction of the word--vectors when building the word embedding.
We then defined new query words specific to each cancer type and repeated the procedure described in \ref{subsection:prostate}. The full list of used query words for each cancer type can be found in the Supplementary Material (section~\ref{sup:queries}),

In Figure \ref{fig:roc_plots} we report the median ROCs for three different distance metrics: INtERAcT (orange curve), cosine (blue curve) and Euclidean (green curve) metrics. In order to obtain a confidence for the curves using the different pathways considered, we built empirical confidence intervals (CIs). The CIs at level 68\% are reported (one standard deviation from the mean) in Figure \ref{fig:roc_plots}. The CIs are generated performing an empirical bootstrap on the different pathways. For each false positive rate level 5000 values are sampled with replacement from the true positive rate values obtained from the different pathways to generate an empirical distribution and build the intervals.

Finally, we compare the similarity of scores predicted by INtERAcT and STRING by computing the Spearman rank correlation between both sets of scores. The values are shown in Table \ref{tab:correlations}. For all  analyzed pathways,  the correlation is positive with a strongly significant $p$-value (the $p$-values range from $10^{-06}$ to $10^{-48}$). The correlation between the negative logarithm of the $p$-value and the number of publications is 0.87, revealing that the main factor determining the significance of the $p$-value is the number of publications used to build the embedding. 

\begin{table}[htb]
\centering
\begin{tabular}{lrrrr}
\toprule
Pathway &  Correlation &  $p$-value  & Proteins &    Papers         \\
\midrule
KEGG Acute Myeloid Leukemia     &     0.340 &   3.0e-14   &      34 &   34532  \\
KEGG Bladder Cancer             &     0.359 &   2.9e-13   &     30 &   35331  \\
KEGG Chronic Myeloid Leukemia   &     0.337 &   1.8e-07   &        23 &   14247  \\
KEGG Colorectal Cancer          &     0.466 &   2.6e-48   &        48 &  118336 \\
KEGG Glioma                     &     0.256 &   6.6e-10   &       36 &   64712  \\
KEGG Small Cell Lung Cancer     &     0.268 &   2.0e-06   &         28 &   32233   \\
KEGG Non Small Cell Lung Cancer &     0.280 &   9.1e-09   &         31 &   67048    \\
KEGG Pancreatic Cancer          &     0.350 &   2.3e-26   &          47 &   62668    \\
KEGG Prostate Cancer            &     0.312 &   1.7e-42   &          67 &  132357   \\
KEGG Renal Cell Carcinoma       &     0.427 &   1.9e-15   &          30 &   37169    \\
\bottomrule
\end{tabular}
\caption{\label{tab:correlations}
{\bf INtERAcT--STRING rank--correlation on KEGG's cancer pathways.}
The table  reports the Spearman correlation and p-values of INtERAcT predictions and STRING--derived  scores for different KEGG pathways. The number of proteins in each pathway, as well as the number of papers used to build each embedding is also reported. For all analyzed pathways  and cancer types, the correlation is positive and highly significant.}
\end{table}

Our findings suggest that while having a large enough corpus is of paramount importance to obtain robust predictions, the number of publications seems  to play  a moderate role in determining the strength of the association between INtERAcT and STRING scores (see Table~\ref{tab:correlations}).
For instance, the highest correlation value 0.47 is found in colorectal cancer, which has the second highest number of publications used to build the embedding. However, prostate--cancer only shows a moderate correlation of 0.31, while having the largest number of publications used. We hypothesize that while having a large corpus of publications is beneficial to build a high--quality embedding, very active fields of research where a high number of publications are available may also be prone to having a high rate of {\it noisy publications}. Here noise can take the form of low-quality publications that report inconsistent results, or studies based on high-throughput analyses with a high false discovery rate. We also note that in taking STRING as ground truth we are implicitly absorbing its false and true discovery rates into our error rates. For instance, interactions reported by STRING that might occur in a different context but not in cancer (e.g. mouse interactions not occurring in cancer) will get penalized as false negatives if INtERAcT correctly predicts them as a non--interaction.

Taken all together and within the  limitation of not having an unbiased ground truth to evaluate our predictions, INtERAcT shows a good agreement with the information reported by STRING. Our results indicate that  our unsupervised approach  is able to recapitulate to a large extent  the knowledge obtained through manual curation of scientific literature.

\section{Conclusions}

We have presented a fully unsupervised method to automatically extract context--specific molecular interaction networks from freely available publications, without any doubt, the fastest growing source of scientific information. Our approach does not require time--consuming manual curation nor labelling of the text. Indeed, no annotations or other manual processing step are required. Furthermore, the results presented here have been obtained without optimization of hyper--parameters.

We have described the steps to reconstruct a context--specific pathway from prostate-cancer publications. When comparing the inferred interactions to STRING, our method outperforms other scores built on commonly used metrics (cosine and Euclidean metric). On a more extensive validation on multiple cancer pathways, the results remain consistent and we have a significant agreement on the information reported by STRING. We would like to highlight that STRING predicts interactions using a combined score that integrates information from many disparate data sources including genomic proximity, gene fusion events, phylogenetic co–occurrences, homology, co–expression, experimental evidence of interaction, simultaneous annotation in databases and automatic text--mining. Text--mining is done using a combination of co-occurrences and natural language processing based on a rule-based system~\cite{saric}. Our methodology on the other hand is a completely unsupervised approach only based on publications that does not require either expert--knowledge  or rules setting. When focusing on reconstructing a prostate--cancer pathway, we achieved a 0.70 AUC score using STRING as benchmark. We notice that the choice of benchmark is likely overpenalising the evaluation of the precision and recall of our method, as STRING reports many interactions that are not cancer-specific.  

We expect the proposed algorithm to be highly relevant for a variety of state of the art text--mining methods. Especially, we are convinced that the proposed methodology can be used to generate hypotheses for detection of biological processes relevant to common and complex diseases and can establish a novel, unsupervised and high--throughput approach to drive drug discovery and advance the frontier of targeted therapies. 

\section{Methods}
\label{sec:methods}

In this section we present the elements that constitute INtERAcT  and describe the approach adopted to automatically build a network of molecular interactions starting from a domain-specific text corpus.
 
\subsection{Text processing}
\label{sec:processing}

We begin by using a basic and lightweight pipeline for text processing. First, we  filter out non--informative words such as extremely common words (e.g. a, the, and other stop--words), rare words (low occurrence in the corpus), non--informative characters like punctuation or isolated numbers and convert text to lower--case. Please, notice that we only remove isolated numbers in order to leave intact and be able  differentiate gene names (e.g. AKT1, AKT2 and AKT3). We next identify  bi--grams -- sequences of 2 words that often appear together and thus are considered a single entity, e.g. New\_York --  by summing up the occurrences of two words  appearing sequentially together in the corpus and setting a  threshold of the minimal number of occurrences. The names of a gene, its aliases and corresponding protein  are treated as synonyms and  mapped to a single name entity  using a dictionary obtained from UniProt\footnote{\url{ftp://ftp.uniprot.org/pub/databases/uniprot/current_release/knowledgebase/idmapping/by_organism/HUMAN_9606_idmapping.dat.gz}}. Sentences are  generated using an English language tokenizer~\cite{bird2009natural} -- a software used to segment a text into linguistic units, in our case, sentences  -- before punctuation is removed. The result of this process is a corpus of sentences  that can be used for further analysis.

\subsection{Word embeddings}
\label{sec:word_embeddings}

\begin{figure}[!ht]
    \centering
    \includegraphics[width=\columnwidth,
    clip]{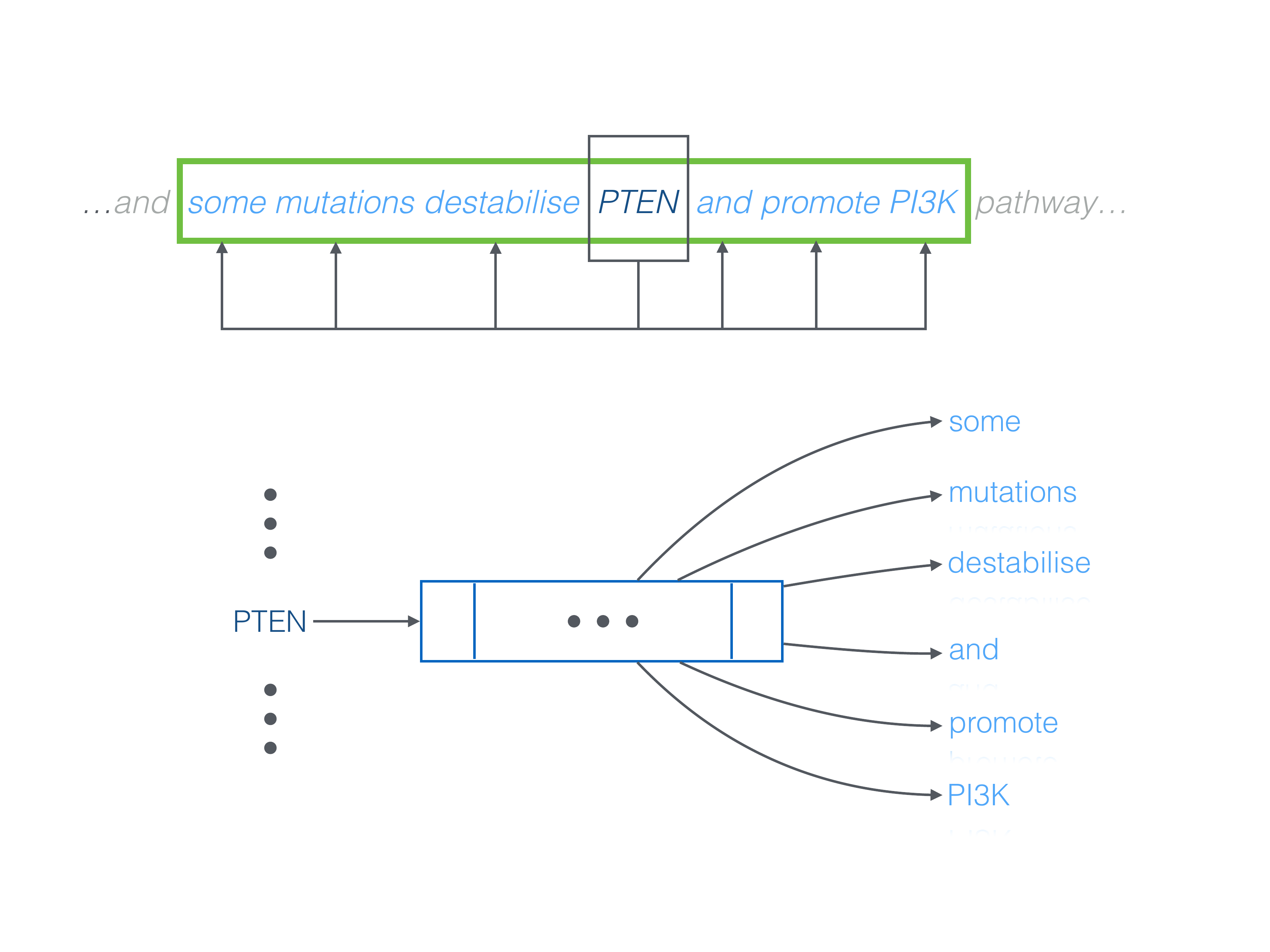}\\
    \caption{{\bf Description of the skip-gram model} used in Word2Vec to find an optimal representation to predict the surrounding context of a target  word.
    The example highlights the window around PTEN, a gene implicated in many cancer processes. The target word, PTEN, is linked to each of its neighboring words and the pairs are fed into the network. The learning process optimizes the probability of predicting the contextual words of  PTEN. }
    \label{fig:skip_gram}
\end{figure}

Word embeddings are the output of a family of methods that produce, starting from raw text, a real vector representation of each word and phrase contained in the original text corpus.
In this work we build the embedding using the  Word2Vec implementation proposed by Mikolov et al.~\cite{mikolov_distributed_2013}, a shallow, two--layer neural network based on a skip--gram model. Briefly, the skip--gram model aims to predict the surrounding words, i.e. the context, of a target word given as an input (see Fig. \ref{fig:skip_gram}). In practice, a word's context is defined by considering a window of size $2n$ to the left and the right of each target word. Each pair target--context word is then   fed into the neural network with a single hidden layer of dimension $d$ that is trained to optimize the  probability of predicting  context words given a  target word as input.  It has been reported that  the quality of the word embedding increases with the dimensionality of the internal layer that produces the vector representation, $d$, until  it reaches a critical point where marginal gain  diminishes~\cite{mikolov_efficient_2013}. Hence this parameter has to be appropriately chosen according to the size of the vocabulary and text corpus.

The word embedding learning process is  naturally optimized to capture the contextual associations between words: If two words tend to appear in similar contexts, they will be mapped into similar word--vectors. In practice, it has been shown that word embeddings outperform  methods based on  counting co-occurrences of words on  a  wide range of lexical semantic tasks and across
many parameter settings~\cite{baroni}.

\subsection{Extracting interactions from the embedding}

Once the embedding is built, our aim is to design a methodology that can predict PPIs based on the distribution of word--vectors in the word embedding. We exploit the idea that molecular entities that interact with each other and are  involved in similar biological processes are likely to appear in similar word contexts, and thus will be  mapped to neighboring positions in the word--vector space. It is hence possible to predict functional similarities between molecular entities based on their mapping and proximity in the word embedding. 

Our task is therefore to find optimal ways of measuring proximity in the word embedding. A first, obvious approach to define proximity between two word--vectors is to use the Euclidean distance and a distance threshold: molecular entities within this threshold can be considered similar and thus predicted to interact.  However, the use of the Euclidean metric, and more generically, the use of $L_k$ norms, is problematic as the high dimensionality of the space can make certain regions of the space too sparse. 
In addition, in high dimensional spaces $L_k$ norms map points to uniform distances from each other, and hence the concepts of proximity, distance or nearest neighbor are not quantitatively meaningful~\cite{aggarwal_surprising_2001}.

INtERAcT exploits  an alternative approach that does not rely on the direct use of $L_k$ norms, but  instead defines similarities between words by looking at the semantic meaning of the neighbors. Specifically, we predict PPIs by comparing the neighborhoods   of words representing molecular entities. To do so, we first need to cluster the word--vectors of the embedding.

\noindent {\bf Clustering words:}
We start by defining $\mathcal{W}$ as the set of $n$ words present in the embedding $\mathcal{E} \in \mathbb{R}^{n \times d}$ where $d$ is the embedding dimension, which corresponds to the dimension of the neural network's hidden layer used to build the embedding.
We  cluster the word--vectors in the embedding space using a K--means algorithm with $C$ clusters. The number of clusters is chosen according to the vocabulary size in order to have both a fine grained word representation and  sufficient number of words per cluster.
Each word is hence associated with a cluster according to the following mapping: 
\begin{equation}
CL : \mathcal{W} \rightarrow \{1, \dots, C\}
\end{equation}.

The obtained clusters group together words that are close in the vector representation space and hence tend to appear in similar contexts in the  corpus. These clusters can then be used   to build  fingerprints  of each entity in the embedding and to convey the semantic meaning of a word based on the cluster membership of its neighbors. 

\noindent {\bf Finding nearest neighbors:}
In order to build word fingerprints, our algorithm requires the identification of the nearest neighbors of each target word. An efficient method to retrieve the topological neighbors without having to compute all pairwise distances at each query is $k$--d trees, a space--partitioning data structure that can be used to organize points in a $k$-dimensional space~\cite{bentley_multidimensional_1975}. A nearest--neighbor--search can then associate every word in the embedding with a set $\mathcal{N}$ of   $K$ nearest neighbors in the embedding:

\begin{equation}
KNN : \mathcal{W} \rightarrow \mathcal{N} \; .
\end{equation}

The optimal number of neighbors depends on the number of clusters $C$, and it is chosen as a trade--off between the benefit of having enough cluster assignment variability among the neighbors, while keeping the neighborhood of each word small when compared to the total word count of the embedding. 
The mapping $KNN$  can be used to efficiently retrieve the shortest paths between two words and identify their nearest words. 

\noindent {\bf Word distribution:}
We are now able to associate each word in the embedding with a discrete probability distribution that can be computed by analyzing the cluster membership of the nearest neighbors.  The number and cluster occupancy of the neighbors can then be interpreted as a discrete probability distribution conveying the semantic meaning of each target word. Furthermore, pair--wise comparisons of these distributions enable us to define similarities between words (see Fig.~\ref{fig:interact_graphical_abstract}).

The pseudo-code used by the described algorithm can be found in the Supplementary Material, section \ref{sup:word_distributions}. The output of the algorithm is  a matrix of probability distributions $\mathcal{D} \in \mathbb{R}^{n \times C}$ where each row contains the cluster assignments of each target word.

\noindent{\bf Computing similarity scores:}
We can finally compute the functional association between  words of interest by computing the similarities between the neighbors' cluster assignments of protein entities in the embedding. We use a score based on the Jensen--Shannon divergence (JSD),  defined as follows:
\begin{equation}
JSD(P || Q) = \dfrac{1}{2}D_{KL}(P || M) + \dfrac{1}{2}D_{KL}(Q || M)
\end{equation}
where $M = \dfrac{1}{2}(P + Q)$ and $D_{KL}$ is the Kullback--Leibler divergence for discrete probability distributions:
\begin{equation}
D_{KL}(P || Q) =  \sum_i P(i) \log \left( \dfrac{P(i)}{Q(i)} \right)
\end{equation}.
The choice of the JSD as a scoring function is motivated by its useful properties. In addition to providing a symmetrized version of the Kullback--Leibler divergence, JSD is  a finite value comprised in the interval $[0, \log(2)]$ \cite{jensen_shannon}, the lowest bound being reached when  two distributions are identical. Furthermore, the square root of the Jensen–Shannon divergence is a metric~\cite{endres_new_2003}, and thus JSD is an appropriate function to capture  similarities between distributions. 

Here we take advantage of the non--negativity of JSD to define the  similarity $S_{ij}$ between words $i$ and $j$  as follows:
\begin{equation}
S_{ij} = \exp( -\alpha JSD_{ij} + \beta)
\label{eq:similarity_score}
\end{equation}
where $JSD_{ij} = JSD(\mathcal{D}_i || \mathcal{D}_j)$ and $\alpha$ and $\beta$ are a scaling and an offset parameters respectively.  In the following, we set the offset parameter $\beta = 0$. Under the transformation defined by Eq.~\ref{eq:similarity_score}, two identical distributions have a score equal to 1 and substantially different distributions (with a divergence close to the $JSD$ upper bound) have a score  $\sim 0.0$. While larger values of $\alpha$ can bring this theoretical minimal value closer to 0, a very high $\alpha$  will make $S_{ij}$ decay too steeply, shrinking the regime where $S_{ij}$ can effectively rank pairs of words  according to their similarity. We found that the choice of   $\alpha=7.5$ and $\beta=0.0$ was empirically efficient at capturing similarities between words  given the theoretical bounds for the $JSD$ (see Supplementary Material Figure \ref{fig:interact_score_description}).

Equipped with the similarity score as defined in Eq.~\ref{eq:similarity_score}, we are now in a position to build a weighted  interaction graph where the nodes are the chosen entities (proteins in our case) and the edges are weighted by the similarity value of the nodes they connect. 

\section*{Declarations}

\subsection*{Acknowledgments}

The project leading to this application has received funding from the European Union’s Horizon 2020 research and innovation programme under grant agreement No 668858. We would like to acknowledge Dr. Costas Bekas and Dr. Yves Ineichen for the useful discussions.

\subsection*{Authors contributions}

M.M., R.M. and M.R.M. conceived the study and analyses. M.M. and R.M. implemented INtERAcT and performed data analysis. M.R.M. provided  biological analysis and interpretation. 
M.M., R.M. and M.R.M. wrote the manuscript with input from all authors. 

\subsection*{Competing financial interests} 

The authors declare no competing financial interest.

\subsection*{Availability of data and materials}

INtERAcT is currently available as a service hosted on IBM Cloud at the following address: \url{https://sysbio.uk-south.containers.mybluemix.net/interact/}. The web service builds a molecular interaction network given word vectors in Word2Vec binary format and a list of molecular entities (example data are made available through a download link in the app).
The article abstracts used to generate INtERAcT protein--protein interaction scores can be freely downloaded from PubMed Central. The article collection as well as access to INtERAcT are also available from the corresponding author on request. 
For this project, STRING interaction scores were downloaded on 19/10/17. STRING historical data can be downloaded from \url{https://string-db.org/cgi/access.pl?footer_active_subpage=archive}, or  obtained from the corresponding author on request.
The networks generated for KEGG cancer pathways and the corresponding word vectors and entities lists can be downloaded at the following link \url{https://ibm.box.com/s/njtqm1alrkq1dxm5jtdnqm8vei4uo58m}.
All the other data generated or analysed during this study are included in this published article and its supplementary information files.

% ArXiv bibliography style
\bibliographystyle{plainnat}
\setlength{\bibsep}{0pt plus 0.3ex}

\begin{footnotesize}
\bibliography{main}
\end{footnotesize}
\newpage

\beginsupplement

\begin{Large}
\noindent \textbf{INtERAcT: Interaction Network Inference from Vector Representations of Words}
\end{Large}\\
\begin{footnotesize}
\noindent Matteo Manica\textsuperscript{1,2,*}, Roland Mathis\textsuperscript{1,*}, María Rodríguez Martínez\textsuperscript{1, $\dagger$}\\
\noindent \textit{ \{tte, lth, mrm\}@zurich.ibm.com}\\
\textsuperscript{1} IBM Research Zürich\\
\textsuperscript{2}	ETH - Zürich\\ 
\textsuperscript{*} Shared first authorship \\
\textsuperscript{ $\dagger$} Corresponding author
\end{footnotesize}

\section*{Supplementary information}

\subsection{Word distributions}
\label{sup:word_distributions}

\begin{algorithm}
\caption{Word distributions}
\begin{algorithmic}[1]
\Procedure{WordDistributions}{$\mathcal{W}, CL, KNN$}
\State $\mathcal{D} \gets \{\}$\Comment{Define a matrix to store distributions}
\For{$w \in \mathcal{W}$}\Comment{For all the words}
\State $D \gets []$\Comment{Define a vector to store $w$ neighbors cluster}
\State $NE \gets KNN(w)$\Comment{Getting K neighbors}
\For{$ne \in NE$}
\State append $CL(ne)$ to $D$
\EndFor
\State $H \gets histogram(D)$
\State append row $H$ to $\mathcal{D}$
\EndFor
\State \textbf{return} $\mathcal{D}$
\EndProcedure
\end{algorithmic}
\end{algorithm}

\subsection{Score analysis}
\label{sup:score_analysis}

\begin{figure}[!ht]
    \centering
    \includegraphics[width=0.75\columnwidth,
    clip]{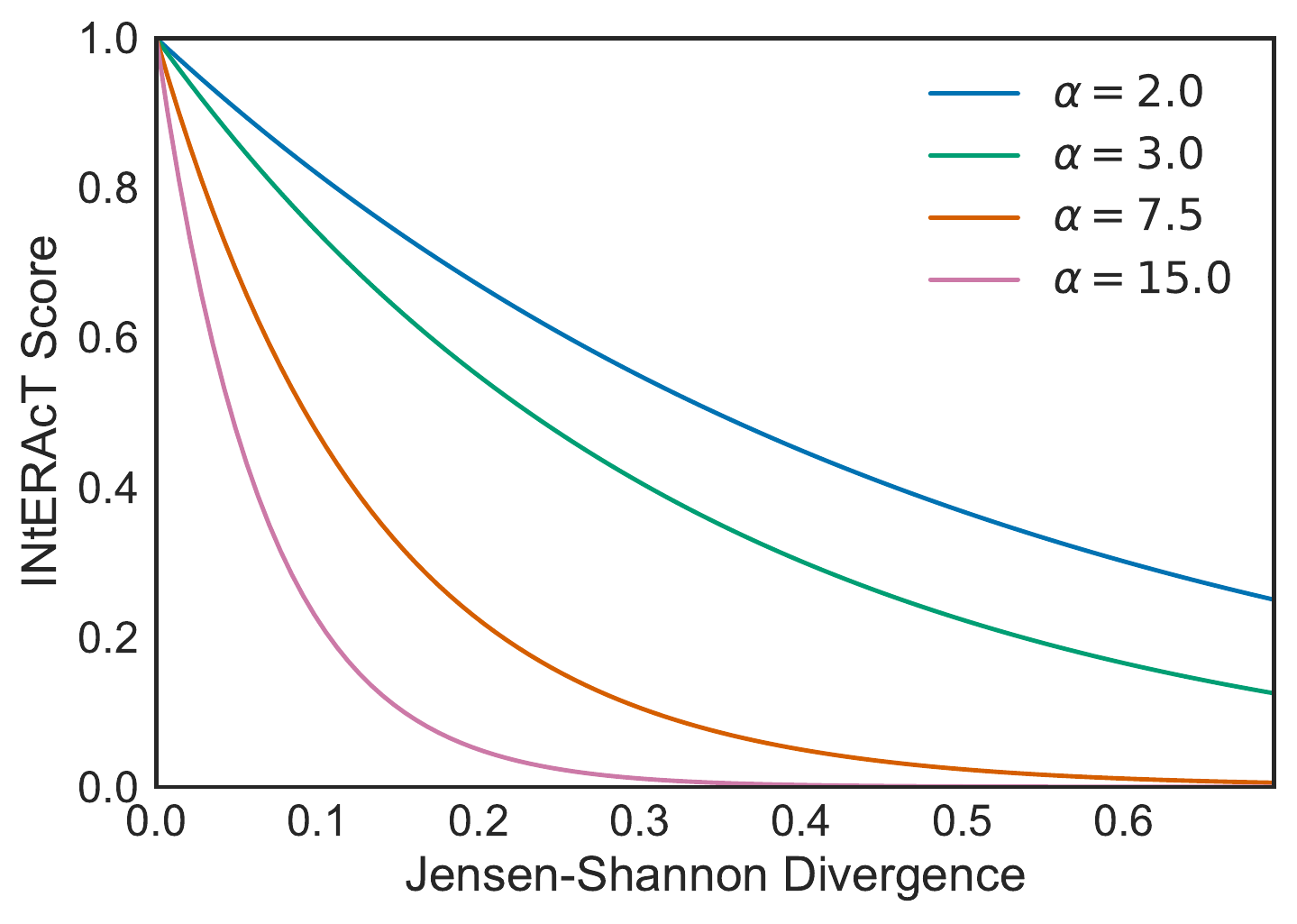}\\
    \caption{\label{fig:interact_score_description}
    {\bf INtERAcT score analysis.} The curves reported describe how the divergence values are mapped into scores by Eq.~\ref{eq:similarity_score} setting $\beta=0.0$ and for different $\alpha$ values. The orange line corresponds to the selected value of $\alpha=7.5$. Other $\alpha$ values don't map properly the divergence values in a [0,1] interval.
    } 
\end{figure}

\subsection{Prostate--cancer scores}
\label{sup:scores}

\begin{table}[H]
\begin{minipage}[t]{0.48\linewidth}
\centering
\begin{tabular}{rrr}
\toprule
Protein &  Protein & Score \\
\midrule
MAPK1 & MAPK3 & 0.87 \\ \hline 
AKT1 & MAP2K1 & 0.83 \\ \hline 
MAP2K1 & MAPK1 & 0.80 \\ \hline 
E2F1 & FOXO1 & 0.80 \\ \hline 
AKT1 & MTOR & 0.80 \\ \hline 
CREBBP & EP300 & 0.78 \\ \hline 
AKT2 & AKT3 & 0.78 \\ \hline 
NFKB1 & RELA & 0.78 \\ \hline 
MAP2K1 & MAPK3 & 0.77 \\ \hline 
E2F1 & E2F2 & 0.77 \\ \hline 
CDKN1A & CDKN1B & 0.77 \\ \hline 
CDKN1B & PTEN & 0.77 \\ \hline 
MAP2K1 & RAF1 & 0.77 \\ \hline 
CCND1 & CCNE1 & 0.76 \\ \hline 
AKT1 & PIK3R1 & 0.76 \\ \hline 
AKT1 & PIK3CD & 0.75 \\ \hline 
CCNE1 & E2F2 & 0.75 \\ \hline 
MDM2 & TP53 & 0.75 \\ \hline 
E2F2 & E2F3 & 0.75 \\ \hline 
MAP2K1 & PIK3R1 & 0.75 \\ \hline 
AKT1 & RAF1 & 0.74 \\ \hline 
CCNE2 & E2F2 & 0.74 \\ \hline 
CCND1 & CDKN1A & 0.74 \\ \hline 
EGFR & IGF1R & 0.74 \\ \hline 
CDK2 & CDKN1A & 0.74 \\ \hline 
\bottomrule
\end{tabular}
\end{minipage}
\begin{minipage}[t]{0.48\linewidth}
\centering
\begin{tabular}{rrr}
\toprule
Protein &  Protein & Score \\
\midrule
CCND1 & CDKN1B & 0.74 \\ \hline
CCNE1 & CCNE2 & 0.74 \\ \hline 
PDGFB & PDGFD & 0.74 \\ \hline 
CDKN1B & FOXO1 & 0.73 \\ \hline 
CCND1 & FOXO1 & 0.73 \\ \hline 
PIK3CD & PIK3R1 & 0.73 \\ \hline 
PIK3R2 & RAF1 & 0.72 \\ \hline 
CDKN1B & E2F1 & 0.72 \\ \hline 
MTOR & PIK3CD & 0.72 \\ \hline 
CDKN1A & E2F1 & 0.72 \\ \hline 
MTOR & PIK3R1 & 0.72 \\ \hline 
AKT1 & AKT3 & 0.72 \\ \hline 
KRAS & NRAS & 0.72 \\ \hline 
CDKN1A & FOXO1 & 0.71 \\ \hline 
CCNE2 & E2F1 & 0.71 \\ \hline 
AKT3 & RAF1 & 0.71 \\ \hline 
AKT1 & MAPK1 & 0.71 \\ \hline 
CDK2 & CDKN1B & 0.71 \\ \hline 
CCND1 & E2F2 & 0.71 \\ \hline 
PTEN & RB1 & 0.71 \\ \hline 
AKT3 & MTOR & 0.70 \\ \hline 
E2F1 & PTEN & 0.70 \\ \hline 
AKT1 & MAPK3 & 0.70 \\ \hline 
CCND1 & CCNE2 & 0.70 \\ \hline 
CDKN1A & TP53 & 0.70 \\ \hline 
\bottomrule
\end{tabular}
\end{minipage}
\caption{\label{sup:prostate_scores}
{\bf INtERAcT top--50 scores for KEGG prostate--cancer pathway.} Top--50 interactions predicted from KEGG prostate--cancer pathway using INtERAcT corresponding to the edges of the graph shown in Figure \ref{fig:prostate_cancer_network_interact}.}
\end{table}

\subsection{PubMed Search Queries}
\label{sup:queries}

\begin{table}[!htb]
\centering
\begin{tabular}{ll}
\toprule
{} &  Query on PubMed \\
\midrule
KEGG Acute Myeloid Leukemia     &    "acute myeloid leukemia"   \\
KEGG Bladder Cancer             &    "bladder cancer" \\
KEGG Chronic Myeloid Leukemia   &    "chronic myeloid leukemia" \\
KEGG Colorectal Cancer          &    "colorectal cancer" \\
KEGG Glioma                     &    "glioma" \\
KEGG Small Cell Lung Cancer     &    "small cell lung cancer" \\
KEGG Non Small Cell Lung Cancer &    "non small cell lung cancer" \\
KEGG Pancreatic Cancer          &    "pancreatic cancer" \\
KEGG Prostate Cancer            &    "prostate cancer" \\
KEGG Renal Cell Carcinoma       &    "renal cell carcinoma" \\
\bottomrule
\end{tabular}
\caption{\label{tab:queries}
{\bf PubMed Search queries for KEGG's cancer pathways.}
In the Table we report the search query that was used for each KEGG cancer pathway. We used the quotation marks to increase specificity. }
\end{table}

\end{document}